\documentclass[sigconf,nonacm]{acmart}

\AtBeginDocument{%
  }
\usepackage{tcolorbox} %
\usepackage{geometry} %
\usepackage{subcaption}
\usepackage{xcolor}
\usepackage{multicol}
\usepackage{multirow}
\usepackage{comment}
\usepackage{enumitem}
\usepackage{algorithm}
\usepackage{algpseudocode}
\setlist[itemize]{topsep=2pt, itemsep=1pt, parsep=0pt, partopsep=0pt}
\usepackage{titlesec}
\titlespacing*{\section}{0pt}{6pt}{4pt}
\titlespacing*{\subsection}{0pt}{4pt}{2pt}
\acmDOI{XXXXXXX.XXXXXXX}
\acmConference[GRAIL @ CIKM '26]
{GRAIL 2026: Workshop on Generative, Retrieval-augmented, and Agentic
Intelligence for Personalization}
{November 8, 2026}{Rome, Italy}
\acmISBN{978-1-4503-XXXX-X/2018/06}

\makeatletter\let\@ACM@checkaffil\relax\makeatother
\makeatletter
\AtBeginDocument{%
  \let\ACM@origsection\section
  \let\ACM@origsubsection\subsection
  \let\ACM@origsubsubsection\subsubsection
  \let\ACM@origparagraph\paragraph}
\makeatother

\begin{document}

\title{From Head to Tail: LLM-Augmented Sequential Recommendation and Trade-off Optimization}

\author{Adeep Hande$^{1}$, Shaghayegh Agah$^{1}$, Neeraj Sharma$^{1}$, Karina Halevy$^{2}$, Sardar Hamidian$^{1}$}

\affiliation{%
  $^{1}$Applied AI Research, Comcast \quad
  $^{2}$Department of Computer Science, Carnegie Mellon University
}

\email{\{adeep\_hande2, shaghayegh\_agah, neeraj\_sharma, sardar\_hamidian\}@comcast.com, khalevy@cmu.edu}
\renewcommand{\shortauthors}{Hande et al.}

\begin{abstract}
Recommender systems exhibit popularity bias, often under-serving long-tail and cold-start items with sparse interaction data. We propose a group-specific data augmentation approach that leverages LLMs to generate synthetic interactions, formulated as a constrained optimization problem, maximizing group-specific NDCG @20 and coverage while bounding overall NDCG@20 degradation within an acceptable threshold. Items are partitioned by interaction frequency to find the optimal augmentation strategy for each group. We further study combining the augmentation across groups simultaneously and find that while individual optimal strategies navigate the multi-group augmentation design, they are not simply additive. We further ablate on the LLM-based augmentation itself, comparing LLM scales, augmentation configuration under token constraints, finding that augmentation quality scales with contextual richness, albeit at a proportionally higher cost. Experiments with SASRec demonstrate the effectiveness of the proposed approach, achieving over 100\% and 60\% improvement in group-specific NDCG@20 for extreme and regular long-tail items, respectively, on our industrial dataset, and 40\%, 25\% for MovieLens-32M, while increasing their coverage without significant overall degradation.

\end{abstract}


\settopmatter{printacmref=true} %
\renewcommand\footnotetextcopyrightpermission[1]{}
\acmConference[GRAIL @ CIKM '26]
{GRAIL 2026: Workshop on Generative, Retrieval-augmented, and Agentic Intelligence for Personalization}
{November 8, 2026}{Rome, Italy}

\maketitle
\vspace{-0.35cm}
\section{Introduction}
Collaborative filtering models learn latent item representations from user-item interactions, but perform poorly when items have inherently sparse interaction data \cite{he2017neural, wang2019neural}. Popularity bias creates a feedback loop that systematically suppresses exposure for items with insufficient co-occurrence signals for both of long-tail and cold-start content \cite{10.1145/3340531.3412152, 10.1145/3437963.3441820}. Within our industrial dataset (PRD) and ML-32M \cite{10.1145/3726302.3730328}, the top 15\% of items account for 95\% and 98\% of all interactions, respectively. Baseline SASRec \cite{kang2018sasrec} exhibits 92-95\% drop from head to tail (Fig \ref{group_distribution}), with cold-start items receiving zero exposure.

Traditional approaches such as self-supervised learning \cite{zhou2020s3rec}, contrastive sequence augmentation \cite{xie2022contrastive}, and adaptive negative sampling \cite{chen2022generatingnegativesamplessequential} address the interaction sparsity while operating within the existing interaction distribution without introducing new exposure for underrepresented items. More recently, LLM-based approaches have been introduced within the recommendation paradigm, generating pairwise sequences \cite{wei2024llmrec}, augmenting interaction graphs \cite{zhang-etal-2025-enhancing, 10.1145/3726871}, metadata-based item enrichment \cite{chen2025enhancing}, and serving LLMs as recommenders for cold-start items \cite{sanner2023large}. \citet{kim2024large} align LLM token spaces with item embeddings, while \citet{huang2025large} simulate user item interactions from filtered candidate sets, and \citet{wang2024llms} cluster LLM recommendations. Other optimizations include training an RL policy for efficient user selection in LLM augmentation \cite{subbaraman2025selectinguserhistoriesgenerate} and post-training LLMs for cold-start reasoning \cite{li2025llm}. In-context learning has also been applied to build synthetic interaction matrices for cold-start users \cite{liang2025incontext, che2024new}.

Recent work has focused more on addressing long-tail recommendations, improving collaborative signals via self-distillation with semantic embeddings \cite{10.5555/3737916.3738755}, and retrieval-augmented LLMs \cite{10.1145/3637528.3671901}, both enhancing existing representations. Towards generating synthetic interactions, TADA \cite{dang2026tail} introduces tail-aware augmentation with heuristic sequence updates. Synthetic interactions generated with LLM have proven to serve as a reliable augmentation signal on a scale \cite{glisovic2026data, wang2024llms, sanner2023large}, reinforcing that LLMs possess sufficient reasoning to act as effective data augmenters. However, these approaches either treat the long-tail as a monolith or do not characterize the tradeoff between overall and group-specific ranking quality under augmentation. \citet{wang2025item} underscores the importance of exploring cold-start items, but does not differentiate between popularity tiers. We build on this by introducing a percentile-based grouping strategy over the interaction distribution, augmenting user histories with LLM-generated interactions drawn from group-specific candidate pools, and analyzing resulting trade-off between overall and group-specific ranking quality.

Our contributions are summarized as follows:
\begin{itemize}[leftmargin=*, nosep]
    \item We propose a group-wise targeted augmentation strategy that classifies items by interaction frequency to tailor the augmenting approach for each group, adapting to the dataset's popoularity distribution.
    \item We characterize the trade-off between group-specific and overall metrics by estimating the Pareto frontier and examining their correlated regions, demonstrating that each group induces a distinct trade-off curve determined by its representation in the historical data. This highlights that a single strategy does not generalize across groups and that one-size-fits-all approaches are suboptimal.
    \item We demonstrate that individually optimal per-group strategies do not compose into a globally optimal solution, confirming cross-group augmentation as a multi-objective optimization problem requiring joint calibration.
      \item  We perform a comprehensive ablation study over candidate selection, LLM scale, insertion position, augmentation percentage, and configurations, introducing a stepwise experimentation framework that narrows the design space without exhaustive search.
\end{itemize}

\vspace{-0.1cm}
\section{Methodology}
\subsection{Dataset and Grouping Strategies}

\begin{figure}[!h]
  \centering
      \includegraphics[width=0.98\linewidth]{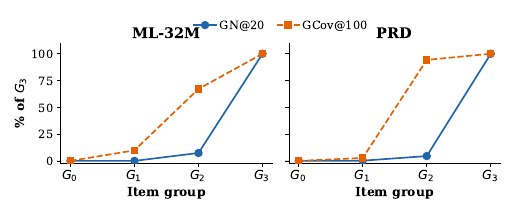}
  \caption{Original Group-wise NDCG@20 (GN@20) by item popularity group on ML-32M (left) and PRD (right) }
  \label{fig:group_ndcg_scatter_plot}
\end{figure}

We evaluate MovieLens-32M (ML-32M)\cite{10.1145/3726302.3730328}, and our \textbf{P}roprietary \textbf{R}eal-world \textbf{D}ataset (PRD) comprising user watch history (TV shows, movies, sports, etc.) from a large-scale entertainment platform. We partition the item catalog in each dataset $\mathcal{I}$ into item-interaction frequency groups at the 75$^{th}$ and 95$^{th}$ percentile computed in log-space over each dataset's interaction distribution. %
We select these boundaries based on the baseline SASRec's performance across the candidate splits, as this partition produces groups with measurably distinct NDCG@20: $G_3$ produce closer to the head, $G_2$ has close to a 95\% drop, while $G_0$/$G_1$ score near-zero performance (Fig \ref{fig:group_ndcg_scatter_plot}).

Unlike fixed threshold approaches that define grouping by arbitrary interaction thresholds, these percentile-based groupings adapt to each dataset's interaction distribution, as the same split yields different absolute thresholds on ML-32M and PRD, reflecting their distinct popularity curves, where $G_1$ (0-75$^{th}$) and $G_2$ (75$^{th}$ - 95$^{th}$) capture extreme and regular long-tail items respectively, while $G_3$ is the popular tier and $G_0$ are cold-start items. %

\begin{table}[!h]
 \centering
 \caption{Group distribution of items (full labeled dataset) and training-data interaction ranges across datasets}
 \resizebox{0.95\linewidth}{!}{
\begin{tabular}{l|l|rc|rc} \hline
\multicolumn{1}{c}{Group} & Percentile & \multicolumn{2}{c}{ML-32M} & \multicolumn{2}{|c}{PRD} \\ \hline
                          &            & Items        & Range       & Items   & Range         \\ \hline
$G_0$                        & N/A        & 71           & 0           & 2,755    & 0             \\
$G_1 $                       & 0-75$^{th}$   & 62,645        & {[}1-23{]}            & 225,806  & {[}1-30{]}    \\
$G_2$                        & 75$^{th}$-95$^{th}$  & 16,414        &    {[}23 - 731{]}         & 60,930   & {[}30-302{]}  \\
$G_3$                        & 95$^{th}$+     & 5,124         &   731+          & 16,671   & 302+ \\  \hline
\end{tabular}}
    \label{group_distribution}
\end{table}

\begin{figure}[!h]
\centering
\includegraphics[height=4cm]{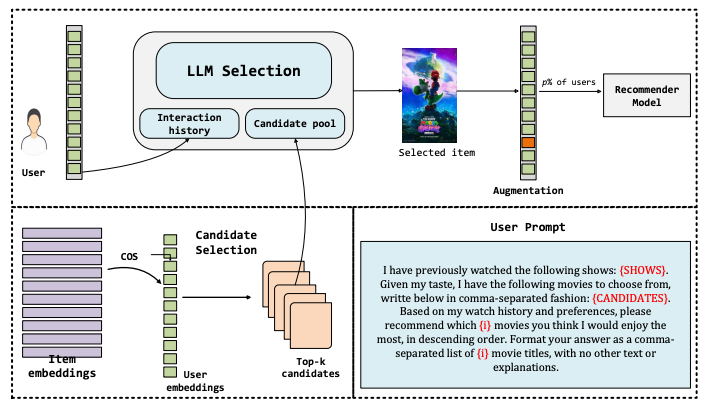}
\caption{Overview of the data augmentation pipeline.}

\label{fig:prompt}

\end{figure}
\subsection{Augmentation Pipeline}
\texttt{Llama-3.3-70B-Instruct} \cite{grattafiori2024llama} serves as the default synthetic interaction generator. For each target group $G_k \in \{G_0, G_1, G_2\}$, the pipeline randomly samples a proportion ($p$) of users and augments each with a single synthetic item ($i=1$) via LLM, selected from among $c=10$ candidates drawn from $G_k$. The augmentation per user is motivated by previous literature \cite{wang2024llms, subbaraman2025selectinguserhistoriesgenerate}, where a single synthetic interaction per user minimizes the risk of shifting the users' preference distribution while still providing a meaningful signal to the base recommender. Candidate selection is performed via cosine similarity between
Linq-Embed-Mistral \cite{LinqAIResearch2024} item embeddings and mean-pooled users' representation on \textbf{PRD}, determined by the encoder ablation in Table \ref{candidate_selection1}. On \textit{ML-32M}, where item metadata is insufficient for meaningful embedding construction, random sampling is used to select candidates for augmentation.
The LLM selects from a closed candidate pool represented in the prompt (Appendix \ref{appendix}), eliminating out-of-catalog selections. The selected item is inserted at $S_{n-2}$, in the user's interaction sequence $S_u=(s_1, s_2, .., s_n)$ immediately preceding the validation ($s_{n-1}$) and test ($s_n$) items. SASRec is retrained from scratch on the augmented interaction matrix. All metrics are averaged over three independent runs. The prompt templates and implementation details are provided in Appendix \ref{appendix}. We formulate evaluation as a constrained optimization, group-specific NDCG@20 (GN@20) and group coverage GCov@100 that measures each group's share of top-100 recommendations per user, are the objectives to maximize, subject to overall NDCG@20 (N@20) degradation remaining within an acceptable threshold.

\subsection{Experiment Designs}
Given the vast search space, optimizing all variables concurrently results in a combinatorial explosion. Therefore, we fixed specific variables such as the quantity and positioning of the augmented data, following prior work \cite{subbaraman2025selectinguserhistoriesgenerate, wang2024llms, liang2025incontext} and prioritized reporting the most impactful axes of variation. To systematically navigate this search space, we utilized a coordinate descent optimization strategy: sequentially fixing hyperparameters to isolate and optimize individual variables step-by-step. This greedy optimization pipeline guarantees practical reproducibility for any dataset and model. Sequential studies designed to isolate the effects of candidate strategy, augmentation proportion, and configuration are discussed below.\\
\textbf{1) Candidate Selection Optimization}: We compare embedding encoders from the MTEB leaderboard \cite{muennighoff2022mteb, enevoldsen2025mmtebmassivemultilingualtext}: \texttt{Linq-Embed-Mistral} (Mistral-7B) \cite{LinqAIResearch2024}, \texttt{qwen3-4b} \cite{yang2025qwen3technicalreport}, \texttt{gte-large-en} \cite{li2023towards, zhang2024mgte} on $G_2$ at $p=5\%$, selecting the encoder that maximizes group-specific NDCG@20. We validate the best encoder against random sampling across several augmentation proportions and quantify the LLM's marginal contribution by comparing paired settings with and without LLM reranking (Table 2 (b)). On ML-32M, we randomly sample the candidates due to lack of sufficient item metadata.\\
\textbf{2) Pareto-front analysis via Augmentation-proportion}: Given the best candidate selection strategy from \textbf{(1)}, we vary $p$ to identify both the group-specific Pareto front, finding the trade-off boundary, and a correlated region where targeted and guardrail metrics improve simultaneously.\\
\textbf{3) Cross Augmentation Optimization}:
Depending on the industry's objectives, optimizing performance across all groups may be necessary rather than focusing on a single specific group. we augment $G_0$, $G_1$, and $G_2$ simultaneously under varying augmentation splits, inspired by the previous step, to evaluate the optimal cross-group augmentation strategy. \\
\textbf{4) LLM Design Space Optimization}: We evaluate whether targeted solutions can be further improved using LLM optimization techniques under a cost constraint. We fix $G_2$ at $p=5\%$, and perform ablation studies over LLM scale (8B, 70B, 405B), insertion position ($S_{n/2}$ vs $S_{n-2}$), user-profiles, and LLM-as-a-judge \cite{fabbri2025evaluating} to isolate each component's individual contribution towards overall and group-wise metrics.  We note that this specific setup serves as an illustrative benchmark to demonstrate the optimization framework, which can be similarly applied to other groups or extended across all groups.

\begin{figure*}[t]
\centering
\resizebox{0.95\linewidth}{!}{
\includegraphics[width=\linewidth, height=8cm]{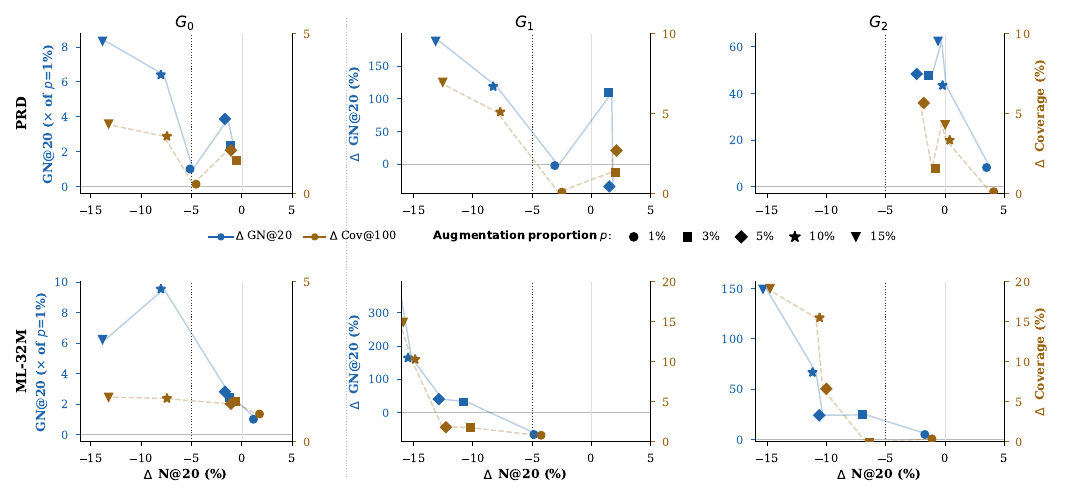}}
\caption{N@20-GN@20 and N@20-GCov@100 tradeoffs across different ($p$). The black dotted line indicates the predefined constraint for the guardrail metric. With a zero baseline for $G_{k}$, improvements are reported relative to 1\% ($p$). All reported numbers are averaged across three runs and are statistically significant.}
\label{fig:dual_axes_tradeoff}
\end{figure*}

\section{Results}
\textbf{1) Candidate Selection Optimization}:
Table \ref{candidate_selection1} (left) compares different embedding-based models for $G_2$ augmentation in PRD dataset at $p=5\%$. The encoder determines the quality of the candidate pool presented to the LLM, providing it with an informed selection space by surfacing the most relevant items with respect to each user. Mistral-7B achieves the strongest balance between the overall and group-specific performance
and consistently improves against random selection at all $p$ values (Table \ref{candidate_selection1} (right)), confirming embedding-based selections improve across all metrics. Paired with Llama-70B (Table-\ref{llm_scale}), we further improve GN@20 by  three times. Random selection degrades both metrics, as sampling from a larger vocabulary produces no semantic relevance to the user. \textit{Mistral-7B} with Llama-70B is used as the default augmentation configuration for all downstream experiments. Embedding-based selection alone yields +5.59\% N@20 and +15.97\% GN@20 (Table \ref{candidate_selection1}), validating that cosine similarity already surfaces relevant candidates. Appending LLM to the pipeline boosts GN@20 to +48.28\%. In addition, the first two rows in Table 3 shows that LLM augmenter is better than random in selecting augmented data.

 \begin{table}[!h]
  \centering
  \caption{Candidate selection optimization (PRD) ($G_2$): (left) encoder-only (right) $\Delta GN@20(\%)$: best encoder vs. random.}
  \begin{subtable}[t]{0.48\columnwidth}
    \centering
    \begin{tabular}{l|cc}
      \toprule
      embeddings & N@20 & GN@20 \\
      \midrule
      mistral-7B  & +6.36 & +15.97\\
      Qwen-3B  & +4.79 & +6.42  \\
      GTE-en  & -6.47& +27.49  \\
      \bottomrule
    \end{tabular}
  \end{subtable}
  \hfill
  \begin{subtable}[t]{0.48\columnwidth}
    \centering
    \begin{tabular}{l|cc}
      \toprule
      $p$ & random & mistral-7b \\
      \midrule
      1\%  & -0.79 & + 8.14  \\
      5\%  &  +24.87 & +48.28  \\
      10\% & +31.15 & +62.42\\
      \bottomrule
    \end{tabular}
  \end{subtable}
  \label{candidate_selection1}
\end{table}
\begin{table}[!h]
\centering
\caption{Comparing two baselines (no augmenters) vs LLM reranking across candidate selection  strategies $G_2$: $p=5\%$.}
\resizebox{0.98\linewidth}{!}{
\begin{tabular}{ll|cc}
\hline
\textbf{Candidate Selection} & \textbf{Augmenter} & \textbf{$\Delta$N@20(\%)} & \textbf{$\Delta$GN@20(\%)} \\
\hline
Random    & -     & -3.44  & -13.04 \\
Random    & Llama-70B          & -4.03      & +24.87      \\
mistral-7B & -          & +5.59  & +15.97 \\
mistral-7B & Llama-70B     & -2.09  & +48.28 \\
\hline
\end{tabular}
}
\label{llm_scale}
\end{table}
\textbf{2)Pareto-front analysis via Augmentation-proportion}:
As shown in Figure \ref{fig:dual_axes_tradeoff}-a, each group traces a distinct tradeoff curve and optimal point governed by both deployment targeted metrics objectives and acceptable degradation in overall performance. We define a guardrail constraint of $\leq 5\%$ \textit{N@20} degradation for identifying viable configurations.
Within PRD, both $G_2$ and $G_1$ show positive correlated
and tradeoff regions. In $G_2$, at $p=1\%$, overall and group NDCG metrics improve by +3.8\% and +8.14\% respectively, targeted metrics continue to rise (GN. of +62\%  and GCov. of +5\% in $p=15\%)$ while N@20 remains within $-2\%$ of the baseline throughout, suggesting that $G_2$ is the most tolerant group to augmentation. Augmenting items that already posses partial learned representations provides SASRec with additional informative signals, allowing the model to learn their representations better, without introducing a sufficient distributional shift that degrades the overall performance. $G_1$ items are extremely sparse in the training data, resulting in under-trained item embeddings sensitive to small volumes of synthetic augmentations. At lower $p$, synthetic interactions can be insufficient to overcome the noise, and their effect becomes volatile and smaller. At higher volumes ($p\geq10\%$), $G_1$ shows consistent group-wise improvement, at the cost of sacrificing the overall performance.

The lack of any N@20 improvements for ML-32M in $G_1$ and $G_2$ plots can be attributed to the candidate selection strategy. For $G_1$, ML-32M evaluates at GN@100 due to limited test users (660) as GN@20 yields near-zero across all configurations. Random candidate selection can introduce noise that prevents joint metrics optimization. Due to the varying distribution of long-tail groups across datasets, the gain in targeted metrics may differ; however, consistent trends are observed across them. For example, given a fixed $p$, $G_1$ shows a larger improvement in the metric than G2 across both datasets, as it starts from a substantially lower baseline. We select $p=3\%$ as $G_1's$ operating point for further cross-group augmentation experiments as it falls in the correlated region while having the lowest augmentation budget to minimize the overall augmentation proportions \cite{dang2026tail}.
\begin{table}[!h]
\caption{\textbf{PRD} Cross augmentation opt. via $p$ distribution.}
\resizebox{0.99\linewidth}{!}{
\centering
\begin{tabular}{r|c|c|c|c} \hline
[$G_0(p)$, $G_1(p)$,$G_2(p)$]  & N@20 & G0=N@20 & G1-N@20 & G2-N@20 \\ \hline
$[5, 3, 15]$   &  -24.49\%    &      $2.96x$    &   +53.95\%       &  \textbf{+45.67\%}     \\
$[5, 5, 5]$ &  -21.38\%   &  $5.74x$         &  +11.31\%       & +19.86\%  \\
$[1.66, 1.66, 1.66]$ & \textbf{-1.06\% }   &  $2.87x$     &  -1.46\%    &  +20.59\% \\
$[1.92, 1.16,1.92]$ & -11.04\% &$2x$  & -7.69\%& +38.41\% \\
$[3.84, 2.32, 3.84]$ & -14.95\%& $4.17x$ & -41.06\%& +17.95\% \\
$[5, 3, 5]$ & -13.8\%  & $5.57x$ & +32.24\%  &  +19.24\%  \\
$[5.77, 3.46, 5.77]$ & -20.6\%& \textbf{$6.61x$} & \textbf{+71.52\%} & 13.44\%\\ \hline
\end{tabular}
}
\label{cross_aug}
\end{table}
 \begin{table}[!h]
 \caption{PRD augmenter ablation performance (\%)}
\centering
\resizebox{0.98\linewidth}{!}{

\begin{tabular}{l|ccc}
\hline

\textbf{Augmenter} & \textbf{$\Delta$N@20(\%)} & \textbf{$\Delta$GN@20(\%)} &  token (x)\\
\hline
Llama-70B-default (baseline) &  -2.09  & \underline{+48.28} & {1.0x}\\
Llama-8B      & -0.15  & +43.45  & {1.0x} \\
Llama-405B    & \underline{+3.11}  & \underline{+48.55} & {1.0x} \\

Llama-70B (early: $S_{n/2}$)   & +0.06  & +31.67  & $1.0x$ \\
LLM-as-a-judge \cite{fabbri2025evaluating}        & \textbf{+6.36} & +34.93 & $2.06x$  \\
Llama-70B + user profiles & -0.16 & +34.09 & $1.57x$ \\
Llama-70B + metadata    & \underline{+3.78}& \textbf{+60.39} & $5.12x$\\
\hline
\end{tabular}
}
\label{augmenter_ablation}
\end{table}

Cold-start items retain their random initialization throughout training, receiving no gradient updates from real interactions, and augmentation provides the only source of learning signal for these embeddings. At $p \leq 5\%$, the augmentation is small enough for the model to improve both targeted metrics while maintaining the N@20 within the acceptable threshold. However, beyond this, the distribution shift steepens the tradeoff sharply. The optimal augmentation is $p=5\%$, balancing a 4$x$ GN@20 and 2$x$ GCov@100 improvement against $\leq 5\%$ N@20 degradation. On ML-32M, the correlated regions can only be observed at $G_0$ for $p=1\%$, owing to significantly smaller candidate pool for augmentation (71 vs 2.7k ).

\textbf{3) Cross Augmentation Optimization}:
We evaluate whether per-group augmentation strategies compose additively.
Naively combining independent optimal proportions (23\% total) degrades N@20 by -24.49\%. This drop may stem from the total augmentation volume itself, their interaction effects, or both. We cap the overall augmentation proportion to $\leq 15\%$ as this represents the maximum evaluated in our per group analysis (Figure \ref{fig:dual_axes_tradeoff}), to keep the guardrail metric decline, N@20 below the certain threshold and measure additive effects in joint optimization.

The non-uniform $p$ allocation across groups and their respective shares of the total augmentation budget is motivated by patterns observed in single-group trade-off optimization.
At 5\% total cross-augmentation with equal splits, both $G_0$ and $G_2$ improve across 3rd and 4th rows, while non-uniform allocation yields a greater overall gain. However, G1 remains sensitive to the splits, improving at higher $p$
that is consistent with our findings that $G_{1}$ requires a minimum augmentation volume to produce stable gains. No configuration simultaneously improves all groups without overall degradation.
Applying group specific $p$ inspired by previous results, last two rows, simultaneously improve all group specific metrics. However, to mitigate degradation in guardrail metrics, further calibration of the joint augmentation is required, framing it as a multi-objective optimization problem.  \\
\textbf{4) LLM Design Space Optimization}
We fix $G_2$ at 5\% augmentation across all ablations for consistency and evaluate against several augmentation strategies as
presented in Table \ref{augmenter_ablation}. Insertion position significantly affects efficacy, with early-augmented diluting their influence as SASRec weighs heavily toward more recent items \cite{oh2024measuring, kang2018sasrec}. User-profile generation \cite{bang2025llm} and llm-as-a-judge \cite{fabbri2025evaluating} adopt a least-to-most prompting paradigm \cite{zhou2022least}, in which intermediate structure are leveraged to iteratively refine the final candidate selection, thereby enhancing overall ranking performance. Augmenting the prompt with additional metadata achieves the highest GN@20, indicating that richer contextual information improves the quality of the resulting augmentations. While these augmentation configurations yield stronger gains, the optimal configuration is itself a tradeoff between guardrail metrics and cost/computational scalability. The base configuration (Mistral-7B + Llama-70B) offers the most favorable cost-quality ratio for production deployment, while metadata enrichment achieves the strongest group-specific improvement at significantly higher computational cost. This motivates task-specific fine-tuning of smaller LLMs, which may yield both stronger augmentation reasoning and lower inference cost compared to prompting a general purpose large language model \cite{li2025llm}.

\section{Conclusion}
We presented a group-wise LLM augmentation framework for sequential recommendation and evaluated it on ML-32M and a large proprietary dataset by separating items into cold-start (G0), extreme long-tail (G1), regular long-tail (G2), and popular tiers (G3). Across candidate selection, augmentation ratio, cross-group splits, and LLM-side design choices, the results show that augmentation is group dependent rather than governed by a single strategy. Embedding-based candidate selection is more reliable than random selection when semantic retrieval is available, and the underrepresented groups differ in how much augmentation they can absorb and in how they trade off with overall NDCG@20. The LLM ablations also show that augmentation quality depends not only on model scale but also on prompting and insertion choices, with clear effects on GN@20, coverage, cost, and guardrail performance. Just as importantly, the best single-group settings do not carry over directly to the joint case, where cross-group augmentation remains a constrained multi-objective problem. A natural next step is to make that joint calibration explicit and test whether the same patterns hold across other recommenders and datasets.

\bibliographystyle{ACM-Reference-Format}
\bibliography{sample-base, software}
\appendix
\section{Appendix}
\label{appendix}
\subsection{Pipeline configuration}
Table \ref{pipeline_config} shows the default configuration used across all experiments unless specified otherwise. All components are interchangeable, and our approach could be substituted with any base recommender, embedding encoder, or instruction-following LLM.
\begin{table}[H]
\small
\centering
\caption{Default pipeline configuration. All components are interchangeable.}
\begin{tabular}{l|l}
\hline
\textbf{Component} & \textbf{Implementation} \\
\hline
Base recommender & SASRec \cite{kang2018sasrec} \\
LLM augmenter & Llama-3.3-70B-Instruct \cite{grattafiori2024llama} \\
Candidate encoder & Linq-Embed-Mistral \cite{LinqAIResearch2024} \\
Candidate selection & Cosine similarity (top-$c$) \\
Items per user & $i=1$ \\
Candidate pool size & $c=10$ \\
Insertion position & $s_{n-2}$ \\
\hline
\end{tabular}
\label{pipeline_config}
\end{table}
\subsection{Percentile-Based Item Grouping}
Algorithm \ref{grouping} describes the grouping strategy. The procedure is agnostic to any dataset and could be applied to any interaction dataset without manual threshold selection. However, the optimal percentile boundaries may vary across domains and recommender architectures, and we recommend validating the split against the baseline group-level performance to ensure distinct tiers (Fig. \ref{fig:group_ndcg_scatter_plot}).
\begin{algorithm}[H]
\small
\caption{Percentile-Based Item Grouping}
\label{grouping}
\begin{algorithmic}[H]
\Require Training set $\mathcal{D}_{train}$, test/val items $\mathcal{I}_{test} \cup \mathcal{I}_{val}$
\Ensure Item groups $\{G_0, G_1, G_2, G_3\}$
\State Count interactions per item: $n_i \leftarrow |\{(u,i) \in \mathcal{D}_{train}\}|$
\State Compute log-counts: $\ell_i \leftarrow \log(n_i + 1)$
\State Remove outliers: exclude items where $|\ell_i - \mu| > 3\sigma$
\State Compute $P_{75}, P_{95}$ over remaining $\ell_i$
\State $G_1 \leftarrow \{i : \ell_i \leq P_{75}\}$
\State $G_2 \leftarrow \{i : P_{75} < \ell_i \leq P_{95}\}$
\State $G_3 \leftarrow \{i : \ell_i > P_{95}\}$
\State $G_0 \leftarrow \{i \in (\mathcal{I}_{test} \cup \mathcal{I}_{val}) : n_i = 0\}$
\end{algorithmic}
\end{algorithm}
$G_0$ items are introduced into training histories via synthetic interactions, thus providing the model with its initial learning signal for these items. The evaluation measures ranking quality on held-out test sequences, and no test interaction is seen during data augmentation step.
\subsection{Prompt Template}
The following listwise prompt template is used for all LLM-based candidate generation experiments. The user's interaction history (\texttt{{SHOWS}}), and candidate items (\texttt{{CANDIDATES}}) are populated per user.
\begin{figure}[h]
\centering
\fbox{\parbox{0.95\linewidth}{\small\itshape
I have previously watched the following shows: \texttt{\{SHOWS\}}.
Given my taste, I have the following movies to choose from, written
below in comma-separated fashion: \texttt{\{CANDIDATES\}}. Based on
my watch history and preferences, please recommend which \texttt{\{i\}}
movies you think I would enjoy the most, in descending order. Format
your answer as a comma-separated list of \texttt{\{i\}} movie titles,
with no other text or explanations.
}}
\caption{Listwise prompt template for LLM-based candidate selection.}
\label{fig:prompt}
\end{figure}

\subsection{Coverage Analysis}
Table \ref{g_coverage} lists the shift in group-wise coverage when each group is augmented at its optimal proportion ($p^*$) from the per-group analysis. The PRD coverage analysis for the selected Pareto-optimal point ({Figure \ref{fig:dual_axes_tradeoff}-b}) shows augmentation redistributes exposure from $G_3$ to the targeted group, increasing item diversity in recommendation outputs. Augmenting $G_1$ in PRD and $G_{0}$ in ML-32M also increases $G_3$ coverage; this happens due to the synthetic interaction that increases both overall and group metrics. Under the $\leq 5\%$ N@20 threshold, ML-32M's $G_1$ and $G_2$, the optimal $p^*$ is limited to $p=1\%$, resulting in negligible coverage shifts.
\begin{table}[H]
\small
\caption{Group Coverage distribution shift for the selected Pareto-optimal ($p$) per Group ($G_{k} (p*)$). Each row shows the coverage shift across all groups when augmenting the target group.}
\begin{tabular}{ll|cccc}
\hline
& $G_k(p^*)$ & $\Delta G_0$ & $\Delta G_1$ & $\Delta G_2$ & $\Delta G_3$ \\
\hline
\multirow{3}{*}{\rotatebox{90}{PRD}}
& $G_0$ (5\%) & +1.35 & -0.38 & +0.40 & -1.38 \\
& $G_1$ (3\%) & 0.0 & +1.35 & -2.77 & +1.43 \\
& $G_2$ (5\%) & 0.0 & +0.46 & +5.68 & -6.14 \\
\hline
\multirow{3}{*}{\rotatebox{90}{ML}}
& $G_0$ (3\%) & +1.26 & -3.92 & -2.78 & +5.43 \\
& $G_1$ (1\%) & 0.0 & +0.81 & -5.16 & +4.35 \\
& $G_2$ (1\%) & 0.0 & -0.90 & +0.36 & +0.54 \\
\hline
\end{tabular}
\label{g_coverage}
\end{table}

\subsection{Data Preprocessing}
For both datasets, we follow SASRec style training and evaluation \cite{kang2018sasrec}, with the penultimate for validation, last interaction for testing, and the rest for training. Interaction sequences are ordered chronologically and truncated or zero-padded to a maximum length of 150. Users with fewer than K interactions are excluded from both datasets.

\end{document}